%% file: scs24paper.tex
\documentclass{scspaperproc}

\usepackage{latexsym}
\usepackage{graphicx}
\usepackage{mathptmx}

\usepackage{amsmath}
\usepackage{amsfonts}
\usepackage{amssymb}
\usepackage{amsbsy}
\usepackage{amsthm}
\usepackage[final]{changes} 

\usepackage[pdftex,colorlinks=true,urlcolor=blue,citecolor=black,anchorcolor=black,linkcolor=black,bookmarks=false]{hyperref}
\input{commands}
\usepackage{hyphenat}
\newtheoremstyle{scsthe}
{8pt}
{8pt}
{\it}
{}
{\bf}
{.}
{.5em}
{}

\theoremstyle{scsthe}

\begin{document}

%
%
\SCSpagesetup{Gil, Ba\c{s}, Jensen, Engelsgaard, Abbiati, and Gomes}

\def\SCSconferencename{Annual Modeling and Simulation Conference}

\def\SCSconferenceacro{ANNSIM'25}

\def\SCSpublicationyear{2025}

\def\SCSconferenceeditors{J.L. Risco-Martín, G. Rabadi, D. Cetinkaya, R. Cárdenas, S. Ferrero-Losada, and A. Bany Abdelnabi}

\def\SCSconferencedates{May 26-29}

\def\SCSconferencevenue{Universidad Complutense de Madrid, Madrid, Spain}

\title{FMI-Based Distributed Co-Simulation with Enhanced Security and Intellectual Property Safeguards}

\author[\authorrefmark{1}]{Santiago Gil}
\author[\authorrefmark{2}]{Ecem E. Ba\c{s}}
\author[\authorrefmark{1}]{Christian D. Jensen}
\author[\authorrefmark{3}]{Sebastian Engelsgaard}
\author[\authorrefmark{1}]{\\Giuseppe Abbiati}
\author[\authorrefmark{1}]{Cláudio Gomes}


\affil[\authorrefmark{1}]{Aarhus University, Aarhus, Denmark}
\affil[\authorrefmark{2}]{R\&D Test Systems, Hinnerup, Denmark}
\affil[\authorrefmark{3}]{Lorc, Munkebo, Denmark}

\maketitle

\section*{Abstract}
Distributed co-simulation plays a key role in enabling collaborative modeling and simulation by different stakeholders while protecting their Intellectual Property (IP).
Although IP protection is provided implicitly by co-simulation, there is no consensus in the guidelines to conduct distributed co-simulation of continuous-time or hybrid systems with no exposure to potential hacking attacks.
We propose an approach for distributed co-simulation on top of UniFMU with enhanced cybersecurity and IP protection mechanisms, ensuring that the connection is initiated by the client and the models and binaries live on trusted platforms.
We showcase the functionality of this approach using two co-simulation demos in four different network settings and \added{analyze the trade-off between IP-protected distribution and} performance efficiency in these settings.

\textbf{Keywords:} Distributed co-simulation, IP protection, collaborative modeling and simulation.

\input{sections/intro}

\input{sections/background}

\input{sections/materials_methods}

\input{sections/results}

\input{sections/conclusion}

\section*{Acknowledgments}
This work has been funded and supported by the DIGIT-Bench project (case no. 640222-497272), funded by the Energy Technology Development and Demonstration Programme (EUDP). 
\added{The authors are grateful to the DIGIT-Bench project partners and the reviewers for the discussions that contributed to this paper.}

\bibliographystyle{scsproc}

\bibliography{references}
\section*{Author Biographies}


\textbf{\uppercase{Santiago Gil}} is a post-doc at Aarhus University, where he completed his PhD in 2024. His research interests include Digital Twins, co-simulation, and Internet of Things. Email address: \email{sgil@ece.au.dk}.

\textbf{\uppercase{Elif E. Ba\c{S}}} holds a PhD in Structural Engineering, with a research focus on hybrid testing. She develops innovative experimental facilities for wind power at R\&D Test Systems. Email address: \email{eeb@rdas.dk}

\textbf{\uppercase{Christian D. Jensen}} is a Professor of Cybersecurity at Aarhus University. He holds a PhD from Université Joseph Fourier. He focuses on trust-based methods and technologies to secure collaboration among entities in open distributed systems. Email address: \email{cdj@ece.au.dk}

\textbf{\uppercase{Sebastian Engelsgaard}} is a Test System Developer at LORC. He specializes in developing and optimizing test systems for offshore renewable energy technologies. Email address: \email{se@lorc.dk}

\textbf{\uppercase{Giuseppe Abbiati}} is an Associate Professor and Head of section at Aarhus University. He leads research on developing DTs for the mechanical and construction industries. Email address: \email{abbiati@cae.au.dk}

\textbf{\uppercase{Cláudio Gomes}} is an Assistant Professor at Aarhus University. He holds a PhD from the University of Antwerp. His research focuses co-simulation and Digital Twins. Email address: \email{claudio.gomes@ece.au.dk}.

\end{document}

%% file: commands.tex
\usepackage{moreverb,url}
\usepackage{amsmath,amssymb,amsfonts}
\usepackage{algorithmic}
\usepackage{graphicx}
\usepackage{textcomp}
\usepackage[table]{xcolor}
\usepackage{orcidlink}
\usepackage{wrapfig,booktabs}
\usepackage{comment} 
\usepackage{array}
\usepackage[T1]{fontenc}
\usepackage{soul}
\usepackage{enumerate}
\usepackage{subcaption}

\usepackage{enumitem}
\usepackage{multirow}
\usepackage{letltxmacro}
\usepackage[most]{tcolorbox}
\usepackage{tikz}
\usepackage{tablefootnote}
\usepackage[nameinlink,noabbrev,capitalise]{cleveref}

\usepackage{listings} 
\usepackage{color}
\usepackage{diagbox}
\usepackage{longtable}

\colorlet{punct}{red!60!black}
\definecolor{background}{HTML}{EEEEEE}
\definecolor{delim}{RGB}{20,105,176}
\colorlet{numb}{magenta!60!black}

\lstdefinelanguage{json}{
    basicstyle=\scriptsize,
    numbers=left,
    numberstyle=\tiny,
    stepnumber=1,
    numbersep=5pt,
    showstringspaces=false,
    breaklines=true,
    frame=lines,
    literate=
     *{0}{{{\color{numb}0}}}{1}
      {1}{{{\color{numb}1}}}{1}
      {2}{{{\color{numb}2}}}{1}
      {3}{{{\color{numb}3}}}{1}
      {4}{{{\color{numb}4}}}{1}
      {5}{{{\color{numb}5}}}{1}
      {6}{{{\color{numb}6}}}{1}
      {7}{{{\color{numb}7}}}{1}
      {8}{{{\color{numb}8}}}{1}
      {9}{{{\color{numb}9}}}{1}
      {:}{{{\color{punct}{:}}}}{1}
      {,}{{{\color{punct}{,}}}}{1}
      {\{}{{{\color{delim}{\{}}}}{1}
      {\}}{{{\color{delim}{\}}}}}{1}
      {[}{{{\color{delim}{[}}}}{1}
      {]}{{{\color{delim}{]}}}}{1},
}
\definecolor{mygreen}{rgb}{0,0.6,0}
\definecolor{mygray}{rgb}{0.5,0.5,0.5}
\definecolor{mymauve}{rgb}{0.58,0,0.82}

\newif\ifcomments %

\newcommand{\claudio}[1]{%
\ifcomments%
  \textit{\textbf{\small{\textcolor{red}{CG: #1}}}}
\else%
\fi%
}

\usepackage{textgreek}

\newif\ifshow 
\showfalse 

\ifshow
  \includecomment{blockcomment}
\else
  \excludecomment{blockcomment} 
\fi

%% file: sections/intro.tex
\section{Introduction}
\label{sec:intro}


The increasing complexity of modern engineering systems has driven the need for collaborative modeling and simulation approaches that enable stakeholders to work together seamlessly.
Distributed co-simulation has emerged to ease the integration and co-development of heterogeneous models and simulations, allowing organizations to contribute their expertise while retaining control over their intellectual property (IP).
However, its adoption is hindered by concerns over cybersecurity and IP protection, especially when a company's model must run on \emph{untrusted platforms} (computers outside the control of the company's IT department) with standalone software.
This problem is recognized in~\cite{Norling2007}, which presents an approach for distributed co-simulation with secure communication between master algorithm and models. However, connections that are initiated from the master (untrusted platform) to the model (trusted platform) pose a security risk that IT departments are often unwilling to take since it requires the trusted platform to have open ports.

Similarly, code executed on untrusted platforms can be extracted or reverse-engineered from memory, as attackers can use debugging tools, memory dumps, or other techniques to analyze and retrieve sensitive data.
This poses a substantial risk to proprietary information, as critical algorithms, configuration parameters, or proprietary logic embedded in the code can be exposed, replicated, or misused by unauthorized parties.

This paper introduces a tool (\added{as an extension} of the UniFMU tool \cite{Legaard2021}) and a methodology for \added{IP-protected distributed co-simulation for collaborative engineering across companies on top of the Functional Mock-up Interface (FMI) standard.}\deleted{\cite{Junghanns2021}}
Unlike conventional distributed co-simulation tools, which often require opening network ports or deploying models on potentially untrusted platforms \added{(see \cref{sec:background}), our approach provides a convenient deployment architecture where connections} are initiated from trusted platforms hosting the models, eliminating the need for exposed ports and thereby reducing susceptibility to cyberattacks.
Moreover, by eliminating the need for models to run on untrusted platforms and keeping them on trusted machines on the client side, our tool enables secure collaboration, particularly in scenarios involving multiple organizations with stringent IP protection requirements, \added{while keeping efficient and adaptable simulation workflows by replacing proxy models with their real counterparts as needed}.

%% file: sections/background.tex
\section{State of the Art}
\label{sec:background}


Co-simulation is a technology that extends the capabilities of simulation, which has been used to answer questions of the type \textit{what-if}\deleted{in combination with modeling~\cite{Maria1997}}, by enabling simulations \deleted{of coupled systems} via the composition of simulation units~\cite{Gomes2018}.
\deleted{Co-simulation is also relevant to hardware-in-the-loop simulations, when some of the simulation units are associated with the I/O of physical equipment.}


Existing standards for co-simulation include FMI, which defines the guidelines for Model Exchange and co-simulation for continuous-time co-simulations;
the IEEE 1516-2010 Standard for Modeling and Simulation (M\&S) High Level Architecture (HLA), which defines the guidelines for distributed (co-)simulations (\textit{Federations}) in discrete-event co-simulations via a common Real-Time Infrastructure (RTI);
the Distributed Co-simulation Protocol (DCP), which provides the guidelines for standardized interoperability on distributed hardware platforms extending from FMI;
and the IEEE 1730-2022 Recommended Practice for Distributed Simulation Engineering and Execution Process (DSEEP), which defines a high-level framework for integrating distributed simulation of lower-level systems.
In this work, we work with the FMI standard.


\begin{table}[!htb]
    \centering
    \scriptsize
    \caption{Summary of approaches in the state of the art.}
    \begin{tabular}{p{.15\textwidth} p{.25\textwidth} p{.38\textwidth} c}
         \hline
         \hline
         \textbf{Contribution} & \textbf{Tool} & \textbf{Communication layer} & \textbf{Standard} \\
         \hline
         Norling et al.~\text{\cite{Norling2007}} & TLM for co-simulation & Stand-alone over TCP & ---\\
         Falcone \& Garro~\text{\cite{Falcone2019}} & HLA Development Kit & HLA Pub/Sub API & FMI \& HLA \\ 
         Skjong et al. \& Sadjina et al.~\text{\cite{Skjong2018,Sadjina2018}} & Coral (now called \textit{The Open Simulation Platform}) & Pub/Sub by slave providers & FMI \\
         Cakmak et al.~\text{\cite{Cakmak2019}} & --- & Models (FMUs) embedded with TCP-based REST APIs & FMI \\
         Schiera et al.~\text{\cite{Schiera2021}} & Mosaik & Master/Slave over TCP (Mosaik Simulator API) & FMI \\
         Meyer et al.~\text{\cite{Meyer2021}} & xMOD & DCP-based over UDP & DCP \\
         Zhao et al.~\text{\cite{Zhao2020}} & LICPIE & Stand-alone over TCP \& UDP & --- \\
         Krammer et al.~\text{\cite{Krammer2021}} & DCPLib & DCP-based over UDP &  DCP \& DSEEP \\   
         Junior et al.~\text{\cite{Junior2016}} & Ptolemy, CertiHLA, \& PyHLA & HLA Pub/Sub API & HLA \\
         Segura et al.~\text{\cite{Segura2023}} & Simulink library & DCP-based & DCP \\
         Reiher \& Hahn~\text{\cite{Reiher2023}} & Portico & HLA Pub/Sub API & HLA \\
         Hong et al.~\text{\cite{Hong2020}} & FMU SDK (now managed by the Modelica Association) & Data Distribution Service Pub/Sub & FMI \\
         Dad et al.~\text{\cite{Dad2021}} & DACCOSIM & ZeroMQ middleware & FMI \\
         Rautenberg et al.~\text{\cite{Rautenberg2023}} & --- & DCP-based over UDP & DCP \\
         Mao et al.~\text{\cite{Mao2023}} & DRLFluent & CORBA & --- \\
         Mehlmann et al.~\text{\cite{Mehlmann2023}} & VILLAS Framework & MQTT, Kafka, FIWARE, WebRTC, RTP, UDP, and TCP & --- \\
         \hline
         \hline
    \end{tabular}
    
    \label{tab:summary}
\end{table}

Existing works in \added{the} literature tackle the need for distributed co-simulation, especially the recent DCP protocol (for continuous-time simulation) and the HLA standard (for discrete-event simulation).
However, there is yet no optimal solution (and guidelines) to conduct distributed co-simulation of continuous-time or hybrid systems which guarantees no exposure to cyber attacks of the machine where the IP-protected models run.
\Cref{tab:summary} shows a summary of contributions in the literature that have addressed techniques to implement distributed co-simulation using different tools, communication methods, and standards.
Of the 16 entries, six \added{work} with FMI, three with HLA, four with DCP, and the rest are not specified.
From these, we can also identify that those following HLA and DCP are more standardized at the communication layer, whereas those following either FMI or none are more diverse.
Although all the approaches provide IP protection implicitly, \added{extra} cybersecurity mechanisms to avoid vulnerability of the machine where the IP is and of the models and binaries \added{include:}~\cite{Falcone2019,Mehlmann2023} ensure that the models and binaries do not need to be shared with third parties,~\cite{Zhao2020,Mao2023} provide session control for simulation units, and~\cite{Norling2007} has encryption in the messaging between the master algorithm and the simulation units, but the connection is initiated from the master, i.e., an untrusted platform.
Our approach also ensures that the models and binaries do not need to be shared with third parties, making the IP hermetic to hostile platforms or users;
additionally, the connection is started from the client-side and, therefore, there is no need for open ports or firewall control.
\claudio{Santiago, I think the small changes made before can just be not highlighted, right?}

%% file: sections/materials_methods.tex
\section{Approach}
\label{sec:materials}
We propose a technique to carry out distributed co-simulation based on the FMI standard.
To do so, we build on top of the existing UniFMU~\cite{Legaard2021}, which already provides the capabilities to split the FMU binary from the model using a Pub/Sub connection over ZeroMQ.
Following the same concept, we split the FMU binary and the model using different processes, which may be executed in different network clients, i.e., one process for the FMU binary and one process for the model execution.

\begin{figure*}[!htb]
    \centering
    \includegraphics[width=0.8\textwidth]{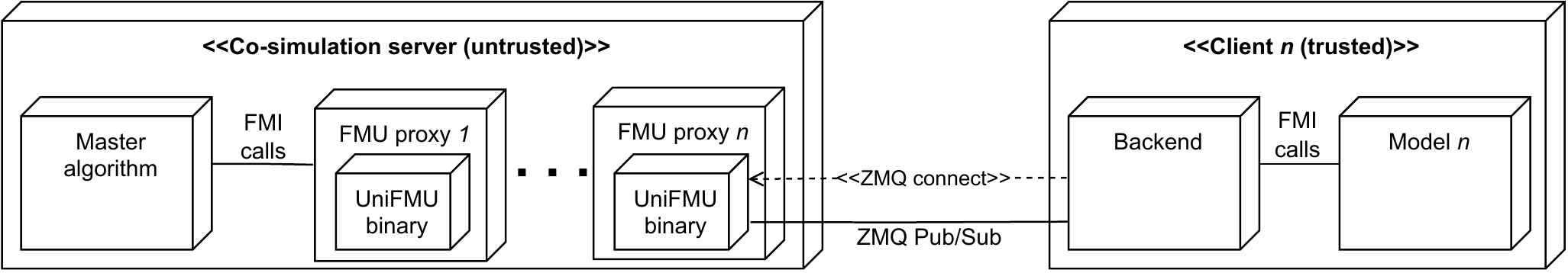}
    \caption{Deployment diagram of distributed co-simulation with UniFMU.}
    \label{fig:architecture_distributed}
\end{figure*}

In this approach, the FMU works as a proxy (without any model), which waits for a connection from the model, through a ZeroMQ connection.
The proxy and the model exchange data in the form of \added{one-to-one acknowledged Protobuf messages following FMI over ZeroMQ, where participants can only access data streams of their own models}.
The proxy forwards the messages from the co-simulation master algorithm to the model and vice-versa, as shown in~\Cref{fig:architecture_distributed}.
\added{The master algorithm, on the other hand, has access to data streams generated from inputs and outputs of the orchestrated proxies and other eventual FMUs.}

When the master algorithm calls the \texttt{instantiate} function on a proxy-model pair, this enters a blocking state, waiting for the client, i.e., the model, to connect to the ZeroMQ broker created by the FMU proxy.
When the model connects to the ZeroMQ broker, the co-simulation starts until termination, releasing the resources of the two processes, that is, proxy and model.
This occurs for each proxy-model pair in the co-simulation.
The master algorithm \added{can combine FMUs whether they are proxies or not.}

It is worth mentioning that the connection is established by the client (the trusted platform), avoiding the need for open ports and firewall control for the connection on the client side, i.e., where the IP lives.
Thus, the open ports are only needed on the proxy side (the untrusted platform), where the collaborative co-simulation is executed, without access to the IP.
At connection time, authentication can also be provided.
This approach can also run distributed co-simulations with existing FMUs, where the box \textit{Model n} in~\Cref{fig:architecture_distributed} is replaced by the existing FMU, and the \textit{FMI calls} are executed on it using \added{the FMPy library}~\cite{fmpy}.


%% file: sections/results.tex
\added{\section{Demonstration and Limitations}}
\label{sec:results}

To demonstrate the use of this approach, we provide \href{https://github.com/INTO-CPS-Association/UniFMU_DistributedCoSimDemo}{two publically available demos on GitHub}~\cite{demounifmu}.
\added{In the first demo (\textbf{Demo 1}), we use the default template created by UniFMU, which provides a pair proxy-model with a set of variables to perform simple calculations of the type sum, rest, string concatenation, and boolean operations; all these operations are executed on demand by the master algorithm, in this case, using FMPy.
In the second demo (\textbf{Demo 2}), we use a co-simulation with three FMUs that replicate a test bench for a 16 MW test facility of wind turbine nacelles. The three FMUs represent the test bench controller, test bench motor, and the drive train generator of the device under test. This demo uses Maestro~\cite{maestro}.
}


For \textbf{Demo 1}, we initialize a co-simulation of the proxy-model pair with FMPy, and iterate through one thousand steps with \added{step size $\Delta_t=0.01$~s} and simple operations, as described in this setting, more precisely, we run one thousand times a set of \texttt{setReal}, \texttt{doStep}, and \texttt{getReal} calls.
For \textbf{Demo 2}, we execute a co-simulation with three proxy-model pairs for a total of $50.0$~s of co-simulation time with a step size of $\Delta_t=0.1$~s.
Once the third FMU pair has established connection, the master algorithm proceeds with the co-simulation execution for $50.0$~s given the co-simulation scenario.
\added{Both demos are executed in real-time and non-real-time modes. For the former, we expect that the average step time ($AS_t$) is $AS_t\simeq \Delta_t$; for the latter (normal simulation time), $AS_t$ depends on the network and computation capabilities, usually $AS_t<\Delta_t$.}

\added{The main limitation is the performance degradation, due to network communication. To understand the trade-off between IP protection and \textit{Performance Efficiency}, we follow the guidelines of the ISO/IEC 25010 for systems and software quality requirements and evaluation.}
For this, we measure the time behavior and resource utilization of the demos in four network settings.
Settings \textbf{1} and \textbf{2} are for collaborative co-simulations where stakeholders can gather physically together, whereas settings \textbf{3} and \textbf{4} can be used for geographically distributed co-simulations, which may compromise security and performance.
\added{Both demos were tested in the four settings, providing the same co-simulation results with different computation times}.
\begin{description}[nosep,leftmargin=0.0em]
    \item[Wireless Local Area Network (WLAN) (Setting 1).] In this setting, we distribute proxy and model processes on two different machines with a private connection over the WLAN using a domestic modem.
    \item[Switched LAN (Setting 2).] In this setting, we use an industrial switch (HP Aruba 2930F) to distribute the two processes as in Setting 1, but this one uses a wired network on the data-link layer instead.
    \item[Public server from a non-controlled private network (Setting 3).] In this setting, we distribute the processes on two different machines, where the proxies are hosted on a public server\deleted{, more specifically,} (an AWS EC2 instance), and the models are hosted on a private machine running on a non-controlled network\deleted{, e.g., at home}.
    \item[Public server from an IT-controlled network (Setting 4).] In this setting, we use the same settings as in \textbf{Setting 3}, but instead, the private machine runs on an IT-controlled network \added{with more strict policies.}\deleted{, where security policies and firewalls are more strict.}
\end{description}

The performance efficiency results for time behavior of this approach are collected in~\Cref{tab:performance_efficiency}.
Notice that the resource utilization for \textbf{Demo 1} is minimal since the computation accounts for the sum of two real numbers and abuses the network resources, whereas the resource utilization for \textbf{Demo 2} is moderate, with the execution of the dynamics of a coupled system.
\added{In particular, for the last row (\textbf{Demo 2} in real-time mode), we clearly see that Settings \textbf{3} and \textbf{4} cannot perform in real-time as expected, exceeding the 50.0s of computation \added{because every step takes longer than required ($AS_t>\Delta_t$)}, opposite to Settings \textbf{1} and \textbf{2}.} 
\added{With~\Cref{tab:performance_efficiency}, we confirm there is a trade-off between hard real-time requirements and (geographical) distribution of IP-protected co-simulation due to incremental network delays.}
If hard or near-hard real-time is needed, following \textbf{Setting 2} is the recommended option, which is the fastest yet secure setup.

\begin{wraptable}{R}{0.6\textwidth}
\scriptsize
\caption{Measures of performance for time behavior (all measures are in seconds).}\label{tab:performance_efficiency}
\begin{tabular}{c c c c c c}\\\toprule  
\textbf{Demo} & \textbf{Real-time mode} & \textbf{Setting 1} & \textbf{Setting 2} & \textbf{Setting 3} & \textbf{Setting 4} \\\midrule
Demo 1 & No & 13.8128 & 1.4747 & 52.4733 & 63.6923 \\
Demo 1 & Yes & 27.2524 & 16.6968 & 61.7995 & 74.5381 \\
Demo 2 & No & 20.8260 & 3.2366 & 79.4714 & 93.6664 \\
Demo 2 & Yes & 50.2971 & 50.1930 & 88.9710 & 93.8596\\  \bottomrule
\end{tabular}
\end{wraptable} 





%% file: sections/conclusion.tex
\section{Concluding Remarks}
\label{sec:conclusion}
In this work, we proposed an approach for distributed co-simulation following the FMI standard \deleted{and using the ZeroMQ Pub/Sub protocol on the existing tool} \added{upon} UniFMU.
\deleted{We demonstrated two different use cases and analyzed the performance efficiency according to the guidelines of the ISO/IEC 25010.}
The demonstration \deleted{and evaluation of the tool} \added{in \Cref{sec:results}} shows that this approach for distributed co-simulation is useful when models need to be hidden from the co-simulation master algorithm, i.e., \deleted{models and their binaries} \added{models} are protected from running on external platforms, ensuring IP protection and integrity.
This is also beneficial since the client hosts the model, avoiding dependencies which may not be contained in the FMUs on the co-simulation server, and decides when to start the connection without the need for open ports or firewall control.
\deleted{The evaluation shows that this tool can be used in different network settings that vary in performance to address case studies involving smaller or larger step sizes and real-time requirements.}


\added{Although the use of strong authentication mechanisms and containerization can improve the security of tools that assume incoming connections on trusted platforms, in practice, engineers participating in these simulations do not necessarily have the required background or permissions to handle this. In our experience with IT departments, changing firewall settings to allow such connections is a cumbersome process.}

Additionally, there are limitations for hard real-time applications with small step sizes due to network delays, especially if hardware-in-the-loop is involved.
In such scenarios, it is suggested to work on a switched LAN with dedicated equipment \added{or using other alternatives, such as DCP co-simulation with operation mode for hard or soft real-time}.
Our approach has still limited support for the FMI 3.0 version, which is a work in progress.
In future work, we plan to improve the authentication and security mechanisms \added{for data confidentiality and integrity} of this approach and evaluate the performance of using Virtual LANs.